\documentclass[reprint,%
]{revtex4-1}

\usepackage{natbib}
\setcitestyle{authoryear,round}
\usepackage[utf8]{inputenc}

\usepackage{braket}
\usepackage{float}      
\usepackage{bbold}      
\usepackage[caption=false]{subfig}
\usepackage{subfig}

\usepackage{bm}         
\usepackage{color}
\usepackage{nicefrac}   
\usepackage{setspace}   
\usepackage{booktabs}   
\usepackage{multirow}
\usepackage{umoline}
\usepackage{amsmath}
\usepackage{amsfonts}
\usepackage{dcolumn}
\usepackage{bm}
\usepackage{graphicx}
\usepackage[dvipsnames]{xcolor}

\newcommand{\kk}{\mathbf{k}}
\newcommand{\rr}{\mathbf{r}}
\newcommand{\Psih}{\hat{\Psi}}
\newcommand{\Psihd}{\hat{\Psi}^\dagger}
\newcommand{\EE}{\mathcal{E}}

\begin{document}

\title{Quantum Fluids of Light \\
\vspace{0.1cm}
\begin{minipage}{0.7\textwidth}
\centering
\textmd{\small{{\it Invited Contribution to} Encyclopedia of Condensed Matter Physics, {\it 2nd edition}}}
\end{minipage}
\vspace{-0.1cm}
}

\author{Iacopo Carusotto}
\email{iacopo.carusotto@unitn.it}
\affiliation{INO-CNR BEC Center and Department of Physics, University of Trento, Via Sommarive 14, I-38123 Trento, Italy}

\date{\today}

\begin{abstract}
In this Chapter, we give a brief review of the state of the art of theoretical and experimental studies of quantum fluids of light. Such systems consist of ensembles of photons that acquire a finite mass from spatial confinement or diffraction and finite binary interactions from the optical nonlinearity of the optical medium. The peculiar properties of these fluids are highlighted in comparison with standard condensed matter systems, with a special emphasis on the novel possibilities that they offer for the generation, the manipulation and the diagnostics of the fluid, as well as on their intrinsically non-equilibrium and/or dynamical nature. Perspectives towards a new generation of experiments on strongly correlated fluids of light and towards opto-electronic applications are finally sketched.
\end{abstract}

\maketitle

\section{Key points/Objectives}
This chapter aims to 
\begin{itemize}
    \item illustrate the general concept of quantum fluid of light and the necessary ingredients
    \item highlight the role of non-equilibrium physics under the effect of driving and dissipation in quantum fluids of light in cavity set-ups 
    \item introduce the $t\leftrightarrow z$ mapping underlying quantum fluids of light in propagating geometries
    \item highlight the exciting perspectives in the direction of realizing strongly correlated fluids of light
    \item briefly summarize how quantum fluids of light can be useful for opto-electronic and quantum applications
\end{itemize}

\section{Introduction}
\label{sec:introduction}

Traditionally, our pictures of matter and light are very different. 

Matter consists of material objects, that can be touched with our fingers and mechanically react to our manipulation in very different ways. For instance, solid matter is rigid; one can take advantage of the weight of liquid matter to float and of its internal friction to swim; complex fluidodynamic phenomena are exploited by birds and aircrafts to fly through gaseous media. At a more sophisticated level, quantum fluids combine hydrodynamics with quantum mechanical features and display a variety of exotic behaviours, from superfluidity in liquid Helium and ultracold atomic clouds~\citep{pitaevskii2016bose}, to superconductivity in electron gases~\citep{Tinkham}, to quantized conductance of incompressible quantum Hall fluids in the presence of strong magnetic fields~\citep{Tong:QHbook}. At completely different scales of energy, exciting properties has been unveiled in the nuclear matter that forms atomic nuclei or neutron stars~\citep{nuclei}. 

On the contrary, light is associated to a flow of electromagnetic waves or, in a quantum description, of corpuscular photons that fly through space at the universal speed $c$ from the emitter to the absorber~\citep{QuantumOptics}. The interesting physics occurs in between, and, in standard configurations, it consists of some in-fly modification of the properties of the beam. Of course this can be extremely rich and diverse, and may involve the interplay of reflection and refraction effects with wave interference phenomena, nonlinear optical processes due to the coupling to electronic transitions in material media, and even quantum effects due to the discrete nature of the photon. Still, all these phenomena are associated to the idea of light has to necessarily keep propagating all the time and can not be stored in some container so to be manipulated as one normally does with matter. 

The concept of {\em Quantum Fluid of Light (QFL)} defies this duality and merges concepts from condensed matter physics and optics to investigate the new collective physics that light displays when it is confined and, then, manipulated as a standard fluid of many interacting photons. This exotic behaviour of course requires very specific conditions to happen: in particular one needs photons to acquire some finite mass and display significant inter-particle interactions~\citep{carusotto2013quantum}.

\begin{figure}
    \centering
    \includegraphics[width=0.9\columnwidth]{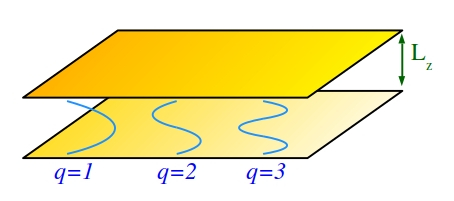}
    \includegraphics[width=0.9\columnwidth]{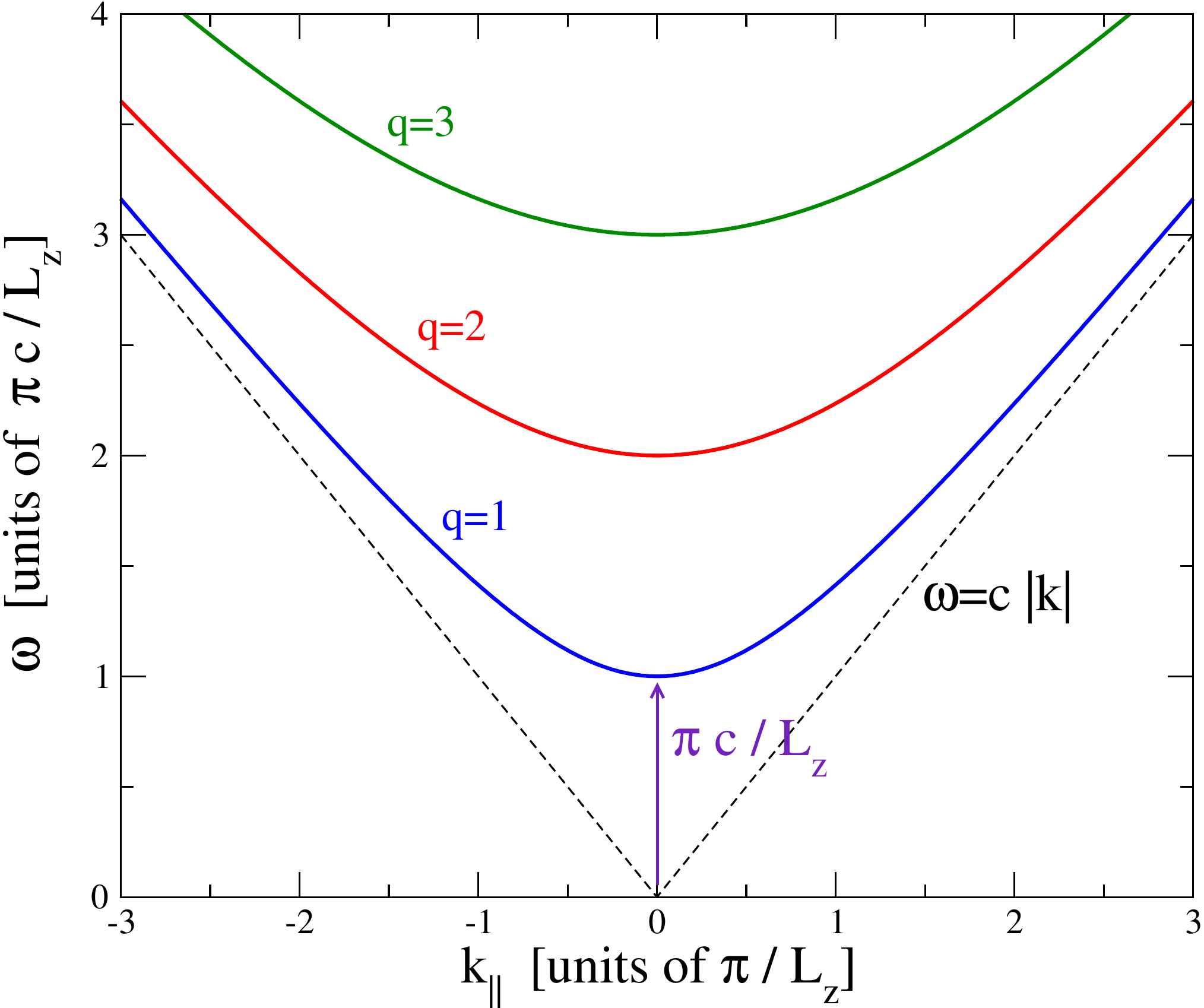}    
    \caption{Effective mass for photons confined in planar microcavities. (Upper panel) Scheme of the device and of the field configuration. (Lower panel) Relativistic in-plane dispersion of photons in the planar cavity.
    \label{fig:basics}}
\end{figure}

From special relativity, we know that massless particles propagate at a fixed speed $c$ and a finite {\em mass} is needed to put particles at rest. Photons are naturally massless particles, but may acquire a finite effective mass when they are confined in space, e.g. in between a pair of plane-parallel metallic mirrors as it is sketched in the top panel of Fig.\ref{fig:basics}. In this simplest configuration, the wavevector along the normal $z$ direction is quantized as $k_z^{(q)}=\pi N / L_z$ in terms of the positive integer number $q$ while the in-plane motion remains free and acquires athemassive relativistic-like dispersion,
\begin{equation}
 \omega^{(q)}(\kk_\parallel)=ck_z^{(q)} + \frac{1}{2k_z^{(q)}}\,\kk_\parallel^2
 \label{eq:dispersion}
\end{equation}
that is plotted in the lower pane of the same Figure. The finite effective mass $m^{(q)}$ arising from the zero-point confinement along $z$ is such that
\begin{equation}
m^{(q)} c^2 = \hbar c k_z^{(q)}
 \label{eq:mass}
\end{equation}
and grows with the quantum number $q$, i.e. the number of nodes of the field in the $z$ direction. As usual, its role is twofold: as a {\em rest mass}, it gives the forbidden energy gap below the dispersion; as the {\em kinetic mass}, it gives the curvature of the dispersion around $\kk_\parallel=0$. A most celebrated family of experiments works with the lowest $q=1$ band of planar semiconductor microcavities enclosed by dielectric mirrors~\citep{carusotto2013quantum}, but a number of other exciting configurations have also been explored with peculiar properties~\citep{schine2016synthetic}. 

\begin{figure}
    \centering
    \includegraphics[width=0.4\columnwidth]{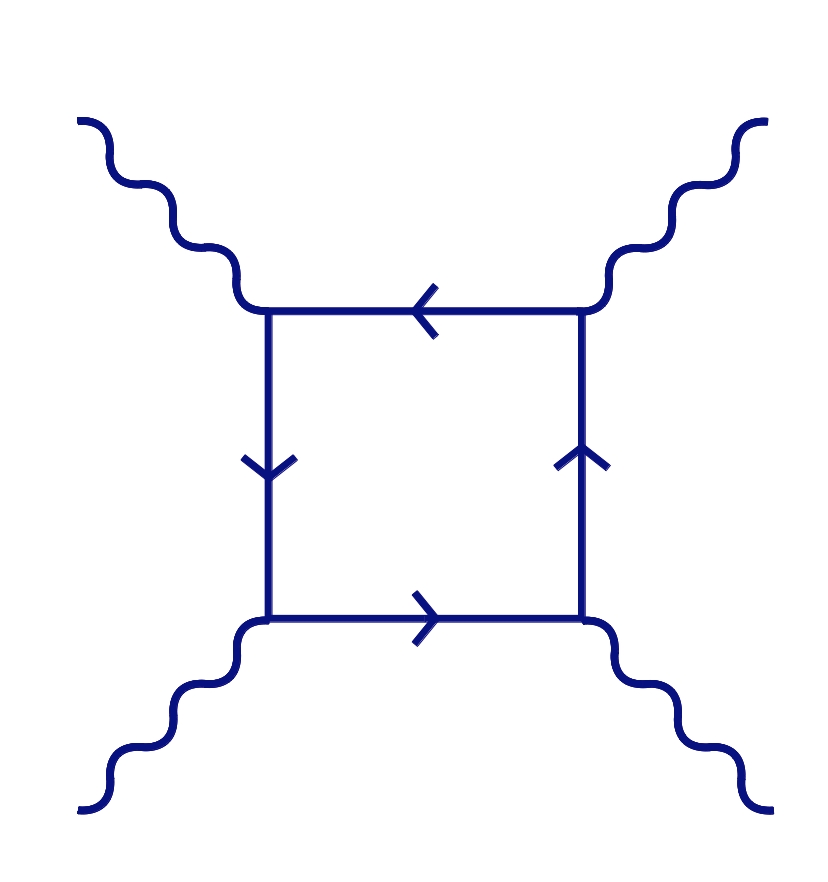}
    \includegraphics[width=0.4\columnwidth]{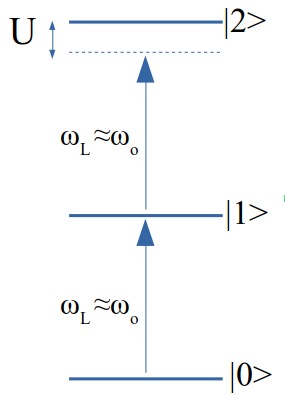}    
    \caption{Basics of photon-photon interactions: (left) Feynman diagram describing interactions mediated by virtual electron-positron pairs in QED; (right) scheme of the energy levels involved in the photon blockade effect in a nonlinear optical cavity.    
    \label{fig:basics2}}
\end{figure}

In contrast to the linearity of Maxwell equations which guarantees a superposition principle for light fields~\citep{jackson1999classical}, photon-photon interactions naturally arise in QED from the coupling of light with the electron-positron quantum field~\citep{heisenberg1936folgerungen}. The simplest Feynman diagram describing photon-photon interactions is sketched in the left panel of Fig.\ref{fig:basics2}, and corresponds to an effective two-photon collision mediated by the creation of virtual electron-positron pairs. In any table-top experiment, the scattering cross-section corresponding to these Heisenberg-Euler processes
\begin{equation}
 \sigma\propto \alpha^4 \, \left(\frac{h}{m_{el} c}\right)^2\,\left(\frac{\hbar \omega}{m_{el} c^2}\right)^6
 \label{eq:H-E}
\end{equation}
is however suppressed not only by the four fine-structure $\alpha=e^2/\hbar c\sim 1/137$ factors stemming from the four interaction vertices and by the small value $h / (m_{el} c)\simeq 2.4\,\textrm{pm}$ of the electron Compton wavelength, but also, and much more dramatically, by the energy denominators corresponding, within perturbation theory, to the virtual intermediate states. For visible or infrared light for which photons have an energy at most in the few eV range, these are dominated by the rest mass of the virtual electron-positron pair which lies in the MeV range and, via the last factor in \eqref{eq:H-E} give a huge $10^{-36}$ suppression factor. As a result, the direct observation of photon-photon scattering in vacuo appears extremely challenging and some experimental evidence was only reported for special configurations based on gamma-rays hitting the ultraintense quasi-bound electric field of a heavy-ion~\citep{atlas2017evidence}. 

The mathematical form of \eqref{eq:H-E} also suggests possible way-outs to this difficulty: at a simplest level of description, the elementary excitations of solid-state insulating materials consist of electrons and holes, i.e. missing electrons in an otherwise filled valence band, which play a very similar role to positrons in vacuo~\citep{ashcroft2022solid}. On the basis of this analogy, it is therefore immediate to expect an enormous reinforcement of the photon-photon scattering cross-section as, in a typical solid-state material, the energy detuning denominator in \eqref{eq:H-E} involves the width of the energy gap in the eV range, and the last factor is thus of order one. This intuitive expectation is fully confirmed by all those nonlinear optical effects that are commonly observed using standard laser sources and can be reinterpreted in terms of photon-photon interaction vertices~\citep{Butcher,Boyd}. Even though this is very oversimplified view and cutting-edge experiments are based on more sophisticated media, still it offers a useful intuition of the underlying physics.
A most efficient strategy to reinforce the optical nonlinearity is to mix the cavity photon mode with some electronic excitation, e.g. an excitonic transition in a quantum well embedded in the cavity layer, and work with the resulting mixed quasi-particles, the so-called exciton-polaritons~\citep{kavokin2011microcavities}.

Further enhancements can be obtained using more sophisticated optical set-ups or engineered meta-materials, so to bring the system into a regime completely at odd with the linear Maxwell equations, where photons behave as impenetrable particles~\citep{chang2014quantum}. After a number of pioneering demonstrations~\citep{Birnbaum:Nature2005}, the most celebrated and promising advances in the direction of strongly interacting fluids of light have been obtained using dressed optical photons in atomic gases in the so-called Rydberg-EIT (Electromagnetically Induced Transparency) regime~\citep{peyronel2012quantum} or microwave photons confined in superconductor-based circuits in the so-called circuit-QED devices~\citep{blais_RMP2021,carusotto2020photonic}. In both cases, the idea is that the refractive index change induced by the optical nonlinearity in the field of a single photon is already sufficient to significantly affect the propagation of the following photons that attempt to penetrate the same spatial region. 

The simplest example of such phenomenon is the so-called {\em photon blockade} effect~\citep{Imamoglu:PRL97}: when a single-mode cavity is filled by a $\chi^{(3)}$ nonlinear medium, the dynamics of the e.m. field can be described as a nonlinear oscillator of Hamiltonian 
\begin{equation}
H=\hbar \omega_o \hat{a}^\dagger \hat{a} + {\hbar \omega_{\rm nl}} \hat{a}^\dagger \hat{a}^\dagger \hat{a} \hat{a}
\end{equation}
whose quantum eigenstates are labelled by the number of photons and have an energy $E_n=n\hbar\omega_o+ \frac{n(n-1)}{2}\hbar \omega_{\rm nl}$ as sketched in the right panel of Fig.\ref{fig:basics2}. Since the separation $E_{n+1}-E_{n}=\hbar\omega_o+(n+1)\hbar\omega_{\rm nl}$ grows with $n$, an incident field tuned to resonance with the $0\rightarrow 1$ transition is detuned from the $1\rightarrow 2$ transition by an effective interaction energy $U=\hbar\omega_{\rm nl}$. As a result, if the nonlinearity exceeds the cavity linewidth $\omega_{\rm nl}\gg \gamma$, the second transition is effectively suppressed and the cavity behaves as an effective two-level system which can only host one photon at a time as if photons were impenetrable objects. 

In the following Sections we are going to review some among the most exciting advances that the concept of Quantum Fluid of Light has given in the last decades and, in particular, of the new fundamental insight that it has given to the physics of fluids in novel non-equilibrium regimes. Sec.\ref{sec:Microcavity} and \ref{sec:Propagating} will be devoted to a brief presentation of the two main conceptual classes of platforms for quantum fluids of light, namely microcavity and propagating geometries. Then, in Sec.\ref{sec:Strong} we will review the on-going research in the direction of strongly correlated fluids of light and topological states of photonic matter. Finally, in Sec.\ref{sec:Conclusion} we will draw some conclusions and we will sketch future perspectives for the field.


\section{Microcavity devices: non-equilibrium physics}
\label{sec:Microcavity}

\begin{figure}
    \centering
    \includegraphics[width=0.99\columnwidth,trim={5cm 1cm 7cm 2cm}]{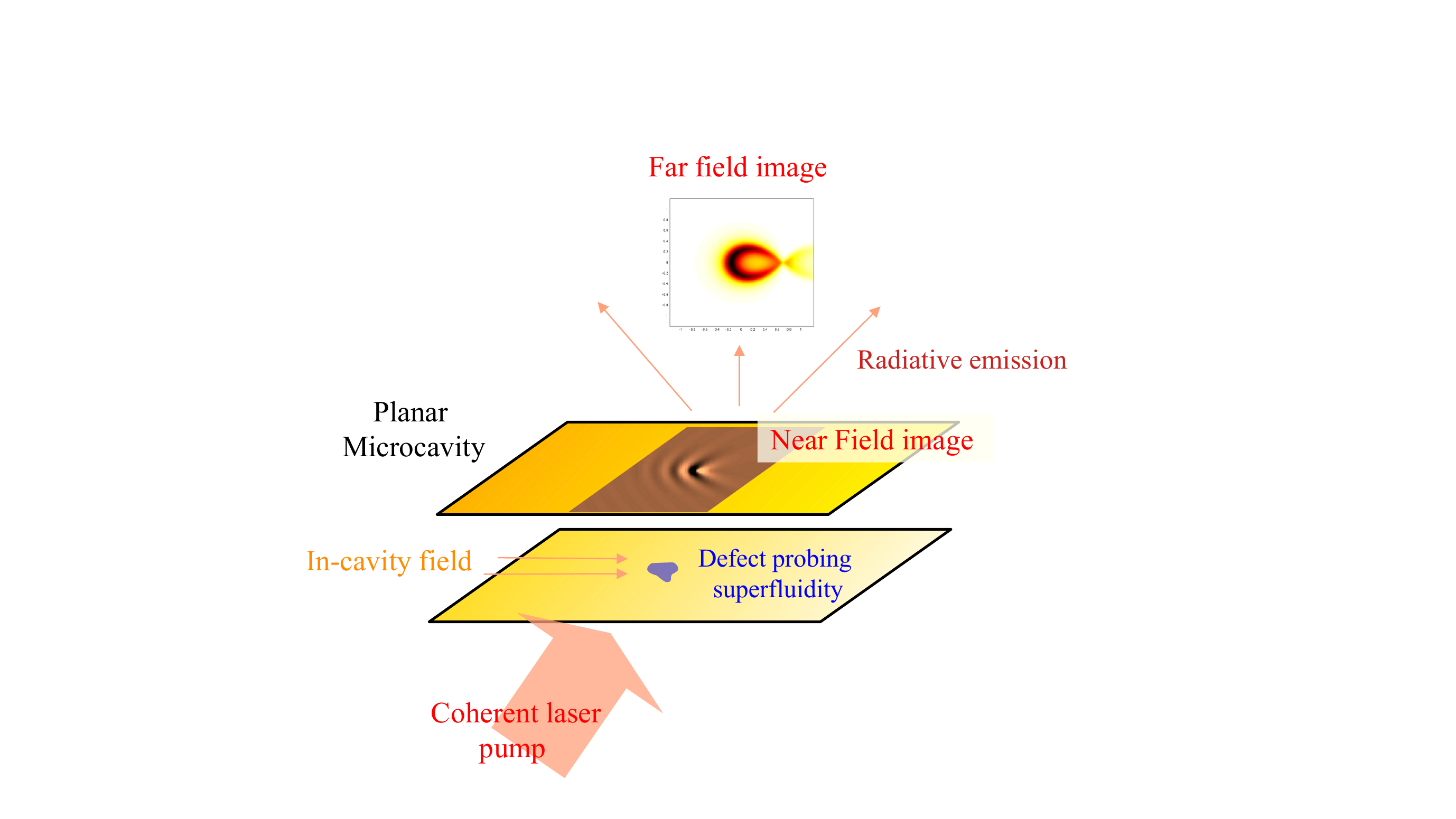}
    \caption{Sketch of a superfluidity experiment with a fluid of light in a planar microcavity geometry. Figure inspired to the theory and experiments in~\citep{Carusotto:PRL2004,Amo:NPhys2009}. For the parameters in the figure, the photon fluid flows faster than the critical speed for superfluidity: a significant intensity modulation appears in the near-field intensity profile (i.e. the spatial density distribution of photons) and a deformed-ring-shaped pattern appears in the far-field one (i.e. the in-plane momentum distribution).}
    \label{fig:microcavity}
\end{figure}

Combining the elements presented in the previous Section, the in-plane dynamics of quantum fluids of light in microcavity configurations turns out to be well captured by a bosonic Hamiltonian of the usual form,
\begin{multline}
H=\int\!d\rr\,\left[\hbar \omega_o \Psihd(\rr)\,\Psih(\rr) + \frac{\hbar^2 }{2m^*} \nabla\Psihd(\rr)\,\nabla\Psih(\rr) + \right.\\ + \left.\frac{\hbar g_{\rm nl}}{2} \Psihd(\rr)\,\Psihd(\rr)\,\Psih(\rr)\,\Psih(\rr)\right]
\label{eq:Hcav}
\end{multline}
whose terms describe, in order, the effective rest energy of the photons, their kinetic energy associated to the kinetic mass, and their (contact) interactions mediated by the optical nonlinearity. This Hamiltonian clearly shows the microscopic processes at play in the quantum fluid of light and their direct equivalence to an ordinary gas of material particles. As such, it is the cornerstone of most theoretical descriptions of the quantum fluid of light~\citep{carusotto2013quantum}.

In addition to the conservative evolution described by the Hamiltonian \eqref{eq:Hcav}, in any realistic configuration one needs to take into account the different loss channels to which photons are naturally subject, in particular radiative losses through the cavity mirrors. In any experiment, the unavoidable radiative losses must therefore be compensated by some continuous pump mechanism to replenish the fluid. This is perhaps the most relevant peculiarity of quantum fluids of light in cavity geometries as compared to standard fluids of material particles and is sketched in Fig.\ref{fig:microcavity}.

Combining the conservative and the driven-dissipative parts of the evolution, the overall dynamics of a weakly interacting fluid of light can be summarized by a generalized driven-dissipative form of Gross-Pitaevskii equation for the expectation value of the Bose field operator $\psi=\langle\hat{\Psi}\rangle$ describing the in-cavity electromagnetic field,
\begin{multline}
 i\frac{\partial \psi}{\partial t}=\left[\omega_o -\frac{\hbar \nabla^2}{2m^*}\right]\,\psi + g_{\rm nl}\,|\psi|^2\,\psi + \\
 +\frac{i}{2} \left[\frac{P}{1+|\psi|^2/n_s}-\gamma
 \right]\psi + E_{\rm inc}(\rr,t)
\end{multline}
where the first and second lines respectively include the conservative evolution and the driven-dissipative terms describing saturable incoherent pumping, losses, and coherent pumping by an external laser field.

In contrast to early belief, the presence of losses is however not just a hindrance and the ensuing driven-dissipative nature of the fluid provides additional experimental knobs. On the one hand, suitably designed incident light fields can in fact be used to inject the photon fluid with the desired spatio-temporal profile and then probe its properties. On the other hand, since all properties of the in-cavity field directly trasfer to the radiated field, the state of the fluid of light can be experimentally reconstructed up to its most subtle quantum correlations just by performing quantum optical measurement on the emitted light~\cite{QuantumOptics}. 

From a wider perspective, these features constitute important assets that enlarge the possibilities of quantum fluids of light well beyond what is normally possible with standard material fluids. Even when the pump and the dissipation establish a dynamical equilibrium, the resulting Non-Equilibrium Steady-State (NESS) is not necessarily thermalized and may display novel features stemming from the non-equilibrium nature. Such a feature was first highlighted in the early '70s when a connection was drawn~\citep{Graham:ZPhys1970} between laser operation and non-equilibrium phase transitions and turned out to play a central role in the numerous studies of non-equilibrium Bose-Einstein condensation effects in fluids of light~\citep{bloch2022non}. Dramatic observable consequences of the non-equilibrium condition have also been observed in different properties of the fluid, from the possibility of condensing into excited states~\citep{Richard:PRL2005,Wertz:NatPhys2010}, to the onset of diffusive rather than sonic collective Goldstone modes~\citep{Wouters:PRA2007,Szymanska:PRL2006}, to subtle features in the definition of superfluid critical velocities~\citep{Wouters:PRL2010}, to peculiar spatio-temporal correlations in the Kardar-Parisi-Zhang universality class~\citep{fontaine2022kardar}, just to quote a few examples. 


\section{Propagating geometries: conservative dynamics}
\label{sec:Propagating}

\begin{figure*}
    \centering
    \includegraphics[width=1.5\columnwidth,trim={1cm 2cm 1cm 6cm}]{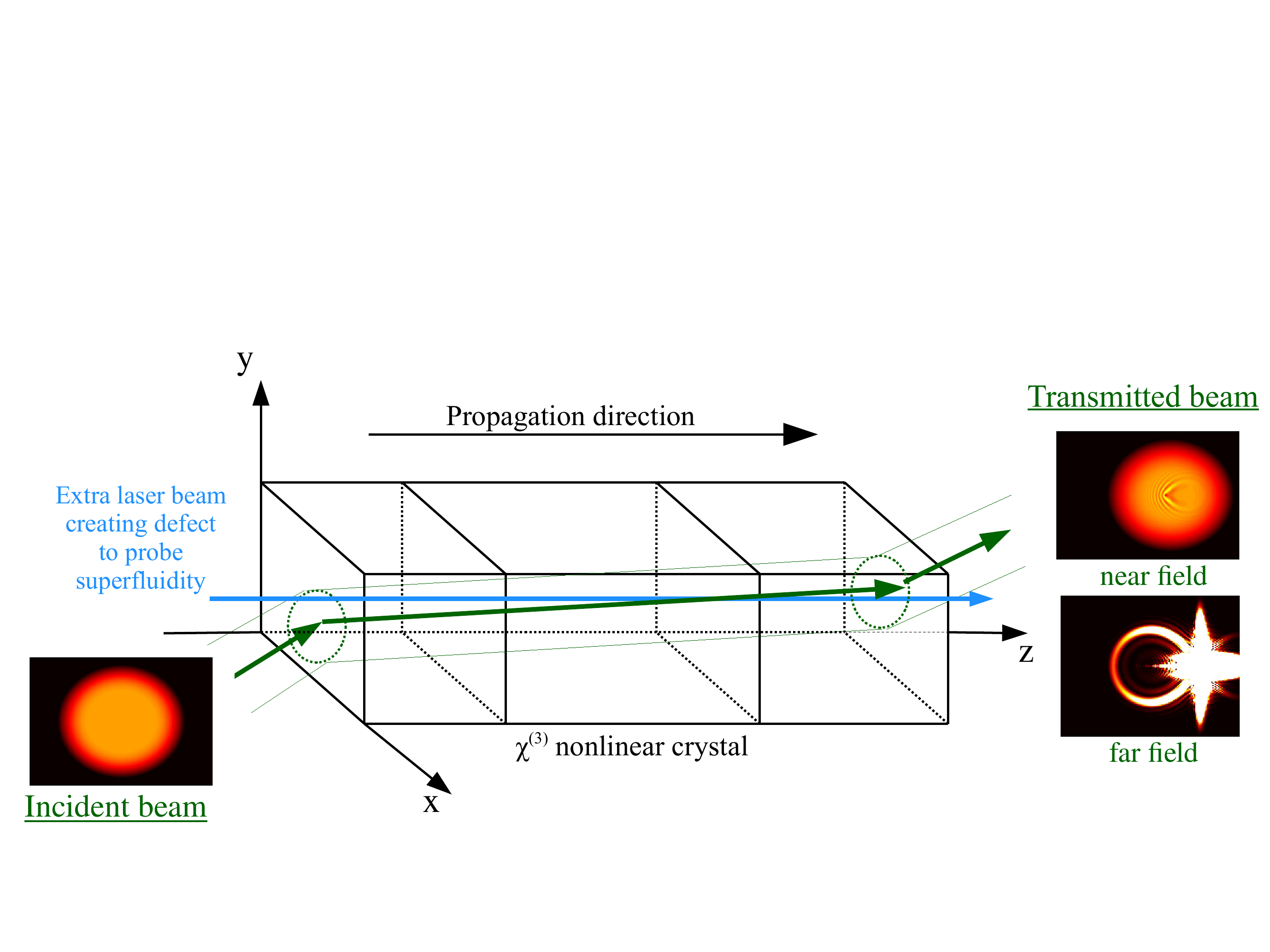}
    \caption{Sketch of a superfluidity experiment with a fluid of light in a propagating geometry. Figure inspired to the theory and experiments in~\citep{larre2015optomechanical,michel2018superfluid}. Also in this figure, the parameters are chosen to have a flow speed faster than the critical speed for superfluidity and clear patterns appear in both the near- and the far-field profiles of the transmitted beam.
    \label{fig:propagating}}
\end{figure*}

A conceptually different class of fluids of light originates from the formal analogy between the paraxial propagation equation for the complex-valued amplitude of monochromatic light in nonlinear bulk media
\begin{equation}
 i\frac{\partial \EE}{\partial z}=
 -\frac{1}{2\beta_0}\nabla_{\perp}^2\EE + V(\rr)\,\EE+ G_{\rm nl}\,|\EE|^2 \EE
 \label{eq:NLSE_z}
\end{equation}
and the Gross-Pitaevskii equation of the dynamics of dilute two-dimensional superfluids: diffraction in the transverse $xy$ plane provides the effective mass term, the transverse refractive index profile induces an external potential $V(\rr)=-\beta_0\,\delta\epsilon(\rr)/(2\epsilon_0)$, and the intensity-dependent dielectric constant  provides photon-photon interactions of strength 
\begin{equation}
G_{\rm nl}= -\frac{\beta_0\,\chi^{(3)}}{2\epsilon_0}\,.
\end{equation}
Here, $\beta_0=\epsilon_0^{1/2}\,\omega_0/c$ is the propagation constant of light at frequency $\omega_0$ in the medium and $\epsilon=\epsilon_0+\chi^{(3)}\,|\EE|^2$ is the nonlinear, intensity-dependent dielectric constant. In this formalism, the slowly-varying complex amplitude $\EE$ is related to the real-valued physical electric field $E$ by
\begin{equation}
E(\rr,t)=\mathcal{E}(\rr)\,e^{i\beta_0 z}\,e^{-i\omega_0 t} + \textrm{c.c.}\,. 
\end{equation}

In the last decades, this configuration has been long used to observe a variety of hydrodynamic phenomena, from optical vortices~\citep{Coullet:1989} to superfluid features~\citep{fontaine2018observation}, to condensation of classical light~\citep{krupa2017spatial,baudin2020rayleigh}.
As a most important conceptual difference from the cavity geometries considered in the previous Section, the role of time is played in these experiments by the propagation coordinate $z$. On the other hand, the use of monochromatic light imposes a rigid $e^{-i\omega t}$ form to the time-dependence of the field.

While this has not been a major concern as long as people were interested in hydrodynamics effects at the mean-field level of \eqref{eq:NLSE_z}, a more sophisticated theoretical model is required to describe quantum phenomena related to the corpuscular nature of the photons. In particular, one needs to develop a theory that is not restricted to monochromatic beams and is able to include a full-fledged temporal dynamics of the field amplitude $\mathcal{E}(\rr,t)$.

A most convenient way to do this is to start from the full classical propagation equation for the spatio-temporal electric field amplitude $\EE(\rr,t)$~\citep{agrawal2001applications,rosanov2002spatial}. Seen from a comoving frame where $\bar{t}=t-z/v_g$, such an equation has again a Gross-Pitaevskii form
\begin{equation}
 i\frac{\partial \EE}{\partial z}=
 -\frac{1}{2\beta_0}\nabla_{\perp}^2\EE + \frac{D_0}{2}\frac{\partial^2 \EE}{\partial \bar{t}^2} + V\,\EE+ G_{\rm nl}\,|\EE|^2 \EE\,,
 \label{eq:NLSE_tz}
\end{equation}
yet in three-dimensions: the propagation distance $z$ plays the role of a temporal coordinate, and the physical time $\bar{t}$ becomes a third spatial coordinate together with the transverse $\mathbf{r}_\parallel=(x,y)$ ones. Here, we have assumed to work with a relatively narrowband signal centered around a frequency $\omega_0$ that propagates at a group velocity $v_g=({\partial k_z}/{\partial \omega})^{-1}$. The effective mass in the temporal direction is determined by the group velocity dispersion $D_0={\partial^2 k_z}/{\partial \omega^2}$. In addition to a spatial dependence, the potential due to the dielectric constant modulation may also display a temporal dependence, $V(\rr,t)$.

This classical field equation can be promoted to a quantum model by means of a non-standard quantization procedure based on the so-called {\em $t\leftrightarrow z$ mapping}~\citep{Lai:PRA1989,Lai:PRA1989b,Larre:PRA2015}. In contrast to the usual canonical quantization where commutation rules are imposed between operators at the same time $t$ and different spatial position $\mathbf{r},\mathbf{r}'$, here the commutation rules are to be imposed between conjugate field operators at the same $z$ position but different rescaled times $\zeta=v_g \bar{t}$ and different positions $\mathbf{r}_\parallel$ along the transverse plane~\citep{Larre:PRA2015},
\begin{equation}
[\hat{\mathcal{E}}(\rr_\parallel,\zeta;z),\hat{\mathcal{E}}^\dagger(\rr'_\parallel,\zeta';z)]=
\frac{2\pi\hbar\omega_0^2\,v_g}{\beta_0\,c^2}
\,\delta^{(2)}(\rr_\parallel-\rr'_\parallel)\,
\delta(\zeta-\zeta')\,.
\end{equation}
To complete the $t\leftrightarrow z$ mapping, we replace the propagation coordinate $z$ with the temporal variable $\tau=z/v_g$ and write a bosonic Hamiltonian in the usual form, where spatial integration is now taken over the three-dimensional space $R=(\mathbf{r}_\parallel,\zeta)$,
\begin{multline}
H=
\frac{\beta_0c^2}{2\pi\omega_0^2}
\int\!d^3R \,\left[\frac{1}{2\beta_0}  \nabla_\parallel\hat{\mathcal{E}}^\dagger\,\nabla_\parallel\hat{\mathcal{E}}-\frac{D_0 v_g^2}{2}\frac{\partial\hat{\mathcal{E}}^\dagger}{\partial \zeta}\frac{\partial\hat{\mathcal{E}}}{\partial \zeta}  \right.+ \\ + \left.V\hat{\mathcal{E}}^\dagger\,\hat{\mathcal{E}}+\frac{G_{\rm nl}}{2} \hat{\mathcal{E}}^\dagger\,\hat{\mathcal{E}}^\dagger\,\hat{\mathcal{E}}\,\hat{\mathcal{E}}\right]\,.
\label{eq:Hprop}
\end{multline}
A quantum version of the paraxial propagation equation \eqref{eq:NLSE_tz} then naturally arises from the Heisenberg motion equation stemming from this Hamiltonian,
\begin{multline}
 i \frac{d\hat{\EE}}{d\tau}=\frac{1}{\hbar}[\hat{\EE},H]=v_g \left[-\frac{1}{2\beta_0}\nabla_{\perp}^2\hat{\EE} + \frac{D_0 v_g^2}{2}\frac{\partial^2 \hat{\EE}}{\partial \zeta^2} + \right. \\ + \left. V\,\hat{\EE}+ G_{\rm nl}\,\hat{\EE}^\dagger\hat{\EE}\hat{\EE}\right] 
\end{multline}
where the $v_g$ factor on the right-hand-side accounts for the $z=v_g \tau$ relation of the  $t\leftrightarrow z$ mapping. 

From the physical point of view, the main difference with respect to the cavity configurations considered in the previous Section is that the field dynamics in propagating geometries is not intrinsically subject to dissipation. Provided the nonlinear medium is transparent on the length scale of the experiment, the light field follows in fact a conservative Hamiltonian evolution similar to the one of cold atomic gases. In contrast to atomic gases, however, the fluid of light is typically found in conditions far from thermal equilibrium condition: as it is sketched in Fig.\ref{fig:propagating} rather than starting from a thermal equilibrium state, the initial condition of the quantum dynamics is imposed by 
the quantum state of the light field incident on the front face of the optical medium. Also in this case, all properties of the final state can be experimentally accessed by means of quantum optical measurements on the transmitted light. 

From a theoretical point of view, this quantum theory of propagating light still relies on heuristic arguments and the $t\leftrightarrow z$ mapping procedure is still awaiting a formal mathematical proof. Yet, first experimental evidence has been reported for a fluid of light subject to a sudden quench of the interaction constant~\citep{steinhauer2022analogue}. This suggests that propagating light is as another promising platform where to study the quantum dynamics of interacting many-body systems.


%

\section{Strongly correlated fluids and topological states of photonic matter}
\label{sec:Strong}

In the previous Sections, we have highlighted a number of exciting hydrodynamic phenomena that take place in weakly interacting fluids of light. Most of these effects are accurately described via classical field equations that generalize the well-known Gross-Pitaevskii equation of dilute Bose-Einstein condensates to the novel context. The new frontier is now to explore regimes where the optical nonlinearities are so important that photons are strongly interacting particles and the fluid develops strong quantum correlations. In this case, a full quantum many-body description is typically required and, in analogy to related work in cold atomic gases~\citep{bloch2008many}, a variety of exotic physical phenomena are expected to arise.

Interesting steps in this direction have been carried out in both microcavity and propagating geometries and closely connect to the on-going research on single-photon nonlinear optics~\citep{chang2014quantum}. A basic building block in view direction is the photon blockade phenomenon already discussed in the introductory section. Beyond the single-mode cavity set-ups in which photon blockade was originally studied~\citep{Imamoglu:PRL97,Birnbaum:Nature2005}, strongly interacting fluids of light are now starting be be investigated both in multi-mode cavity geometries as well as in propagating geometries. 

On the former side, a most remarkable accomplishment has been the experimental realization of a Mott insulator state of impenetrable photons in an array of nonlinear superconductor-based cavities for microwave photons~\citep{ma2019dissipatively}. On the latter side, a marked antibunching has been observed in the transmitted light after propagation through a strongly nonlinear medium formed by dressed atoms in the Rydberg-EIT configuration~\citep{peyronel2012quantum}. For a different level configuration providing attractive rather than repulsive photon-photon interactions, experimental evidence of two-photon bound states was also reported~\citep{firstenberg2013attractive}.

An even richer variety of strongly correlated phases of matter has been anticipated to arise in fluids of light in the presence of synthetic magnetic fields and/or in topological photonic structures~\citep{ozawaRMP2019topological}: in spite of the electric neutrality of the photon, photonic configurations have been in fact explored where light propagation is subject to analogs of the Lorentz force acting on charged particles in magnetic fields. As a most celebrated experimental observation, unidirectional light propagation on chiral edge states has been demonstrated for photonic lattices with topologically non-trivial band structures~\citep{wang2009observation,hafezi2013imaging,rechtsman2013photonic}. More recently, {\em topological laser} devices have been demonstrated in topological lattices endowed with gain~\citep{st2017lasing,Bahari:Science2017,bandres2018topological}: in addition to being a promising device for opto-electronic applications, such systems can be seen as a novel example of quantum fluid of light living in a chirally moving one-dimensional geometry and with peculiar statistical properties~\citep{Amelio:PRX2020}. Further exciting perspectives are offered by the so-called {\em synthetic dimensions}, which allow to realize quantum fluids of light living in high-dimensional geometries beyond the usual three spatial dimensions~\citep{ozawaprice2019topological}.

In combination with strong photon-photon interactions, topological photonic systems have been anticipated to lead to optical analogs of fractional quantum Hall physics~\citep{Umucalilar:PRL2012,Kapit:2014PRX}. Among the most remarkable pioneering experimental results in this direction, we can mention the chiral currents observed in a few-site plaquette for strongly interacting microwave photons~\citep{Roushan:2016NatPhys} and the realization of baby Laughlin liquids in twisted cavities for visible photons~\citep{clark2020observation}. Motivated by these early achievements, a strong activity is presently in progress in the international community towards the realization of macroscopic samples of topoological states of photonic matter~\citep{kurilovich2021stabilizing}. 

\section{Conclusions and Perspectives} 
\label{sec:Conclusion}

The exciting advances that we have summarized in this Chapter provide an overview of the intriguing physics of quantum fluids of light and suggest the promising developments that one may anticipate for this field of research in the next future. After decades where hydrodynamic descriptions of optical systems have been considered as just a mathematical analogy and an intellectual curiosity, the recent advances have demonstrated how the physical picture of light as an ensemble of interacting photons provides a novel example of condensed matter system where to explore, in particular, uncharted lands of non-equilibrium quantum many-body physics. 

A further interest for these systems comes from the straightforward inclusion of quantum fluids of light in devices for opto-electronic applications: the close relationship between lasing and Bose-Einstein condensation has been known for decades~\citep{Graham:ZPhys1970}, but has received a renewed interest in the last years with the development of novel concepts of coherent light sources~\citep{bloch2022non}. On a longer run, an even more ambitious application could be to realize a topological quantum computing architecture~\citep{Nayak:RMP2008} based on topological states of photonic matter.

\bibliography{biblio.bib}

\end{document}